\documentclass[10pt]{wlscirep}
\usepackage{geometry} 
\usepackage{comment}
\usepackage{color}
\usepackage{subfigure}
\usepackage{amsmath,amssymb,amsthm,ulem,amsfonts}
\usepackage{graphicx}				
\usepackage{hyperref}								

\title{Controlling Light Transmission Through Highly Scattering Media Using Semi-Definite Programming as a Phase Retrieval Computation Method}

\author[1,*]{Moussa N'Gom}
\author{Miao-Bin Lien}
\author{Nooshin M. Estakhri}
\author{Theodore B. Norris}
\author{Eric Michielssen}
\author{Raj Rao Nadakuditi}
\affil[1]{Department of Electrical \& Computer Engineering, University of Michigan Ann Arbor, MI 48109}

\affil[*]{Corresponding author: mngom@umich.edu}

\newcommand{\diag}[1]{\mathrm{diag}(#1)}
\DeclareMathOperator*{\argmin}{\arg \min}


\begin{abstract}
Complex Semi-Definite Programming (SDP) is introduced as a novel approach to phase retrieval enabled control of monochromatic light transmission through highly scattering media. In a simple optical setup, a spatial light modulator is used to generate a random sequence of phase-modulated wavefronts, and the resulting intensity speckle patterns in the transmitted light are acquired on a camera. The SDP algorithm allows computation of the complex transmission matrix of the system from this sequence of intensity-only measurements, without need for a reference beam. Once the transmission matrix is determined, optimal wavefronts are computed that focus the incident beam to any position or sequence of positions on the far side of the scattering medium, without the need for any subsequent measurements or wavefront shaping iterations. The number of measurements required and the degree of enhancement of the intensity at focus is determined by the number of pixels controlled by the spatial light modulator.
\end{abstract}
\begin{document}

\maketitle

\noindent 
\section{Introduction}
The ability to focus light through highly scattering  translucent or `opaque' random media has been a long-standing challenge. Strong scattering impedes information transfer through random media and is a limiting factor in many optical imaging and characterization systems. The last decade however has witnessed a tremendous advance in techniques for focusing fields that pass through complex media through shaping of incident wavefronts. We refer the reader to Rotter and Gigan \cite{gigan} which provides an excellent and timely review of the topic.

 Vellekoop and colleagues \cite{vell} were the first to show that by shaping incident wavefronts with a spatial light modulator (SLM) one can compensate for scattering phenomena to produce light focused onto a point inside or beyond the scattering medium \cite{Ivoth, mosk, Ivo}.   Their technique and variants thereof have been used to focus light through white paint layers, eggshells, clouds, dense fogs \cite{yaron}, and biological tissue \cite{yuh}. The ability to focus light at a desired location within or beyond highly scattering media has potential  applications ranging from biomedical engineering  and microscopy to endoscopy \cite{Yu} (and references therein), optical trapping \cite{mazilu}, super resolution imaging \cite{putten, park}, and nano-positioning \cite{putten2}. 

Their enormous potential notwithstanding, widespread use of the above and related schemes in real-world applications has somewhat lagged expectations.  Needless to say, adoption of any wavefront shaping method to focus light through highly scattering media hinges on the practicality of the  scheme for determining the optimal wavefront.  Methods developed to date by and large belong to one of two categories.
\\(i) Iterative methods that use measurements of transmitted field magnitudes (square roots of field intensities). The majority of wavefront shaping methods developed to date use measurements of transmitted field magnitudes to progressively improve field focus by sequentially changing the phase retardation imposed on each pixel of the input beam using an SLM.  These methods maximize intensity of the output field  at a defined location by changing one SLM pixel at a time \cite{ivo2}.  While very powerful, these methods tend to be slow and prone to convergence to a local optimum.  They also must be repeated every time a new target output intensity profile is specified. Oftentimes, speckle correlations can be exploited to refocus fields onto a neighboring pixel with only minor degradation in intensity, without restarting the algorithm. Recent advances have increased the depth of the medium to where the ``memory effect'' (i.e, correlations in the fields) can be reliably exploited \cite{schott}. In the deep medium regime, where the fields are decorrelated and the memory effect no longer holds, refocusing still requires the procedure to be repeated every time a new target output profile is specified.
\\(ii) Non-iterative methods that use measurements of transmitted field magnitudes and phases.  Knowledge of the transmission matrix (TM) of the medium allows for the optimal wavefront to be computed non-iteratively for any desired output.  If  the desired output is a focal spot, then only a portion (e.g. a row or set of rows) or the TM are required.  The first measurement of an optical TM was undertaken by Popoff et al. \cite{popoff}. Using a common-path interferometer equipped with an SLM, the TM of an opaque layer of ZnO nanoparticles was measured and then used to generate optical foci in the output plane. This method was also used to demonstrate image transmission through the opaque layer \cite{popoff2}. Most approaches developed to date to measure TMs rely on holographic methods.  Needless to say, methods for determining TMs that do not require a reference beam would be tremendously useful in this context. A first step in this direction recently was taken by Dr\'emeau et al \cite{angelique} , who used a phase retrieval algorithm to measure the complex TM of a highly scattering medium using a digital micro-mirror.

This work introduces a new waveform-shaping technique for producing focused fields that uses semidefinite programming (SDP) to construct portions of the medium's TM from intensity measurements only; once the relevant parts of the TM are calculated the method proceeds like any of the above non-iterative schemes.   The SDP approach leverages a rigorous yet flexible computational  framework that utilizes the algorithm developed by Waldspurger et al \cite{irene} to retrieve the phase of elements of the TM after recording intensities in the desired focal spot produced by several randomly structured illuminations of the scattering medium.  We note that SDP has been successfully applied to many problems in other fields, ranging from X-ray and crystallography imaging and diffraction imaging to Fourier optics and microscopy \cite{fajwel}.  We believe this work to be the first to apply this powerful algorithm to problems related to optical phase retrieval enabled wavefront control. We demonstrate that SDP allows  the construction of the incident wavefronts that generate intense foci anywhere beyond the medium from a single set of measurements of intensities generated by random illuminations.
\noindent

\section{Algorithm}\label{algorithm}
\subsection{Preliminaries}\label{prelim}

\begin{figure}[htbp]
\begin{center}
\includegraphics[width=3.5in]{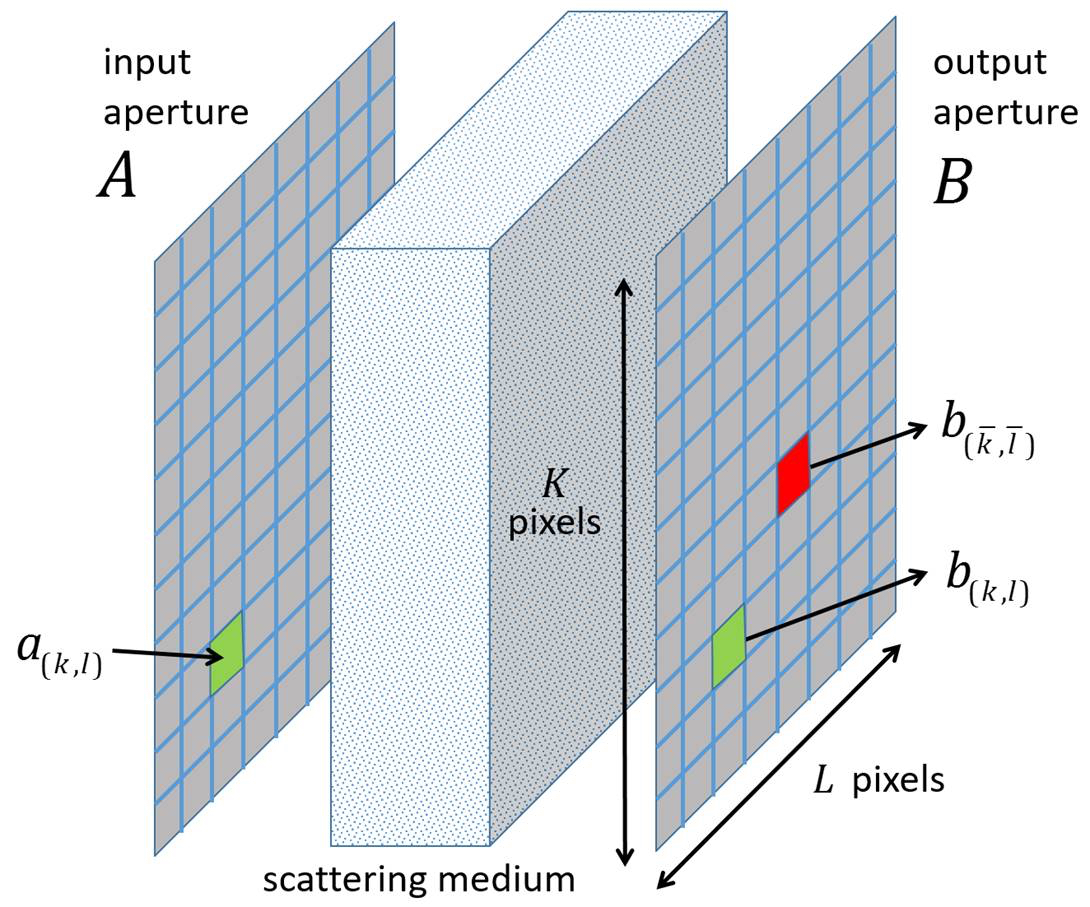}
\caption{Algorithm conceptual setup: the scattering medium is sandwiched between two rectangular apertures (\textit{A} and \textit{B}) . The corresponding  optical setup is shown in Fig. \ref{opt_desg}.}
\label{concept}
\end{center}
\end{figure}

The setup under consideration is shown in Fig. \ref{concept}.  A highly scattering, random medium is sandwiched in between input and output apertures $A$ and $B$.   In what follows,  we assume both apertures are rectangular and comprised of  $M = K \times L$ pixels.  Light in the input aperture impinging on the random medium therefore is characterized by $M$  complex-valued electric field samples $a_{(k,l)}$  with $1 \leq k \leq K, 1 \leq l \leq L$ .   Depending on the experimental setup used, we can control both the amplitude and phase of $a_{(k,l)}$, or as in more common, only its phase.  Likewise, light in the output aperture exiting the random medium is characterized by $M$  complex-valued electric field samples $b_{(k,l)}$  with $1 \leq k \leq K, 1 \leq l \leq L$ .   Our experimental setup only allows for measurements of the magnitude of each $b_{(k,l)}$.  Note: for expositional simplicity we assume that the number of pixels in the input and output apertures are equal.  In an experimental setup, they often are different with the number of pixels in $B$  exceeding those in $A$ . The arguments below however are easily generalized to this setting.   Likewise,  we characterize electric fields in $A$ and $B$  by scalars. Again, the arguments below are easily modified to account for the vector nature of the fields. Alternatively, polarizers can be added to the input and output apertures to effectively scalarize the field. To simplify the notation, we replace all double indices $(k,l)$ by a single index $m = (k-1) K + l $ and immediately ``flatten'' all inherently multidimensional tensors introduced below accordingly.

The fields in $A$ and $B$  are related by the TM of the random medium.  Define the $M \times 1$ vectors
$$\underline{a} = \begin{bmatrix} \cdots & a_{m} & \cdots \end{bmatrix}^{T},$$
and
$$ \underline{b} = \begin{bmatrix} \cdots & b_{m} & \cdots \end{bmatrix}^{T}.$$
The random medium's $M \times M$  TM $\uuline{T}$ relates $\underline{a}$ and $\underline{b}$ as
\begin{equation}\label{Efields}
\underline{b} = \uuline{T} \cdot \underline{a}.
\end{equation}
For future reference, we denote 
\begin{equation}
\uuline{T} = \begin{bmatrix} & \vdots & \\
							 \cdots & T_{m}^{m'} & \cdots \\
                                    & \vdots & \end{bmatrix} 
                                    = \begin{bmatrix} \vdots \\ \underline{z}^{\dagger}_{m} \\ \vdots \end{bmatrix},
\end{equation}
where ``$m$'' denotes the row index of $\uuline{T}$ while  $m'$ denotes the column index of $\uuline{T}$; .  With this notation, the $M \times 1$ vectors 
\begin{equation}
\underline{z}_{m} = \begin{bmatrix} \vdots \\ [T_{m}^{m'}]^* \\ \vdots\end{bmatrix},
\end{equation}
contain the complex conjugates of the elements of row $m$ of the transmission matrix $\uuline{T}$.  Note: use of the terminology ``transmission matrix" implies that the vectors  $\underline{a}$ and $\underline{b}$  resolve all (propagating) modes in  $A$ and $B$ .  In practice, we often violate this condition, especially in  $A$ .  Undersampling of fields in  $A$   however does not invalidate the method described below. 

The objective is to sculpt the incident field so that it produces a maximally focused spot at a predefined location in the output plane subject to an energy constraint.  Mathematically, our goal is to determine the incident field vector $\underline{a}$ that maximizes $|b_{\overline{m}}|$ for a predefined  $\overline{m}$  subject to $||\underline{a}||_{2}^{2} =1$ .  We will consider two cases: (i) full amplitude and phase control of all $a_{m}$ and  (ii) phase-only control of $a_{m}$, \textit{i.e.}, $|a_{m}|^{2} =1/M$ for all $m$.  It follows from (\ref{Efields}) that $b_{\overline{m}}= \underline{z}^{\dagger}_{\overline{m}} \cdot \underline{a}$. Hence, the above objectives would be easily realized if $\underline{z}^{\dagger}_{\overline{m}}$ , i.e. the $\overline{m}$ th row of the transmission matrix $\uuline{T}$ , was known . Unfortunately $\underline{z}^{\dagger}_{\overline{m}}$  is not known.  Below, we first describe a procedure for determining $\underline{z}^{\dagger}_{\overline{m}}$  from measurements of magnitudes of transmitted fields at pixel $\overline{m}$ (Section \ref{ph_ret_alg}).  Expressions for the optimal  $\underline{a}$ follow easily (Section \ref{foc_wave}).

\subsection{Phase retrieval algorithm}\label{ph_ret_alg}

To determine $\underline{z}_{\overline{m}}$ , we illuminate the random medium with $N$  randomly selected or ``trial'' fields $\underline{a}^{(t)}$, $t = 1, \ldots, N$ ,  of uniform magnitude across the input aperture, \textit{i.e.}, $|\underline{a}_{m}^{(t)}|^2 = 1/M$ for all $m$.  We record the magnitude of the field at pixel $\overline{m}$  for all excitations
\begin{equation}
c_{\overline{m}}^{(t)} = |b_{\overline{m}}^{(t)}| = |\underline{z}^{\dagger}_{\overline{m}} \cdot \underline{a}^{(t)}|,
\end{equation}
and write
\begin{equation}
b_{\overline{m}}^{(t)} = c_{\overline{m}}^{(t)} \cdot [u_{\overline{m}}^{(t)}]^{*}\end{equation}
where $u_{\overline{m}}^{(t)} = \exp(-i\, \arg[b_{\overline{m}}^{(t)}])$ contains all of $b^{(t)}_{\overline{m}}$'s phase information. Next, define the $N \times 1$ vectors
\begin{subequations}
\begin{align}
\underline{b}_{\overline{m}} &= \begin{bmatrix} \cdots & b_{\overline{m}}^{(t)} & \cdots \end{bmatrix}^{T} \\
\underline{u}_{\overline{m}} &= \begin{bmatrix} \cdots & u_{\overline{m}}^{(t)} & \cdots \end{bmatrix}^{T},\\
\underline{c}_{\overline{m}} &= \begin{bmatrix} \cdots & c_{\overline{m}}^{(t)} & \cdots \end{bmatrix}^{T},
\end{align}
\end{subequations}
and the $N \times M$ matrix
\begin{equation}
A = \begin{bmatrix} \vdots \\ (\underline{a}^{(t)})^{\dagger} \\ \vdots \end{bmatrix}.
\end{equation}
Note that we have abused notation in thus defining $\underline{b}_{\overline{m}}$ -- we do so to keep our notation light. It follows from Eq. (\ref{Efields})  that
\begin{equation}
\underline{b}_{\overline{m}} = [A \cdot \underline{z}_{\overline{m}}]^{*}.
\end{equation}
Or equivalently, that
\begin{equation}
\diag{\underline{c}_{\overline{m}}}\cdot \underline{u}_{\overline{m}} = A \cdot \underline{z}_{\overline{m}}.
\end{equation}
We therefore proceed to determine $\underline{z}_{\overline{m}}$ and $\underline{u}_{m}$ by solving the following optimization problem:
\begin{equation}\label{eq:phasecut l2}
\begin{gathered}
\textrm{minimize} \,\,|| A \cdot \underline{z}_{\overline{m}} - \diag{\underline{c}_{\overline{m}}}\cdot \underline{u}_{\overline{m}} ||_{2}^{2}, \\
\textrm{subject to } \underline{u}_{\overline{m}} \in \mathbb{C}^{N}, |u^{(i)}_{\overline{m}}| = 1, \, \underline{z}_{\overline{m}}\in \mathbb{C}^{M}.
\end{gathered}
\end{equation}

We solved the above problem using the PhaseCut algorithm described in \cite{irene}, restated below for expositional purposes in terms of the above notation. We start out to note that if $\underline{u}_{\overline{m}}$ were known, then  the vector $\underline{z}_{\overline{m}}$ that minimizes Eq. (\ref{eq:phasecut l2}) can be explicitly computed and is given by
\begin{equation}\label{eq:zfromu}
\underline{z}_{\overline{m}} = A^{\dagger} \cdot \diag{\underline{c}_{\overline{m}}} \cdot \underline{u}_{\overline{m}},
\end{equation}
where $A^{\dagger}$ is the Moore-Penrose pseudoinverse of $A$. Thus, following PhaseCut \cite{irene}, our strategy is to solve Eq. (\ref{eq:phasecut l2}) for $\underline{u}_{\overline{m}}$ and then determine $\underline{z}_{\overline{m}}$ via Eq. (\ref{eq:zfromu}). To that end, we note that plugging in $\underline{z}_{\overline{m}}$ in Eq. (\ref{eq:zfromu}) into the objective function on the right hand side of Eq. (\ref{eq:phasecut l2}) yields 
\begin{align*}\label{eq:phasecut uobjective}
&||A \cdot \underline{z}_{\overline{m}} - \diag{\underline{c}_{\overline{m}}} \cdot \underline{u}_{\overline{m}}||_{2}^{2} \\
& = ||A \cdot A^{\dagger}\cdot \diag{\underline{c}_{\overline{m}}}\underline{u}_{\overline{m}} - \diag{\underline{c}_{\overline{m}}} \cdot \underline{u}_{\overline{m}}||_{2}^{2} \\
& = || (A \cdot A^{\dagger} - I) \cdot \diag{\underline{c}_{\overline{m}}}\underline{u}_{\overline{m}}||_{2}^{2}\\
& = \underline{u}_{\overline{m}}^{\dagger} \diag{\underline{c}_{\overline{m}}} \cdot  (A\cdot A^{\dagger} - I) \cdot (A\cdot A^{\dagger} - I)^{\dagger} \cdot \diag{\underline{c}_{\overline{m}}} \underline{u}_{\overline{m}} \\
& = \underline{u}_{\overline{m}}^{\dagger} \diag{\underline{c}_{\overline{m}}} \cdot (I-A\cdot A^{\dagger}) \cdot \diag{\underline{c}_{\overline{m}}} \underline{u}_{\overline{m}}\\
& = \underline{u}_{\overline{m}}^{\dagger} \diag{\underline{c}_{\overline{m}}}\cdot P \cdot \diag{\underline{c}_{\overline{m}}} \underline{u}_{\overline{m}},\\
& = \underline{u}_{\overline{m}}^{\dagger} Q_{m} \underline{u}_{\overline{m}}
\end{align*}
where we have utilized the fact that $(A\cdot A^{\dagger} - I)^H \cdot (A \cdot A^{\dagger} - I)^{\dagger} = (I - A \cdot A^{\dagger})$, the $N \times N$ matrix $P$ is defined as
$$ P = I - A \cdot A^{\dagger},$$ and the $N \times N$ matrix $Q_{\overline{m}}$ is defined as
$$Q_{\overline{m}} =\diag{\underline{c}_{\overline{m}}} \cdot P \cdot \diag{\underline{c}_{\overline{m}}}.$$
Consequently, for $\underline{z}_{\overline{m}}$ given by Eq. (\ref{eq:zfromu}), the optimization problem involving $\underline{z}_{\overline{m}}$ and $\underline{u}_{m}$ in Eq. (\ref{eq:phasecut l2}) can be expressed as an optimization problem involving only $\underline{u}_{\overline{m}}$ given as:  
\begin{equation}\label{eq:phaseret original}
\begin{gathered}
 \textrm{minimize}\,\, \underline{u}_{\overline{m}}^{\dagger} Q_{\overline{m}} \underline{u}_{m} = \textrm{Tr}(Q_{\overline{m}} u_{\overline{m}}u_{\overline{m}}^{\dagger}), \\
 \textrm{subject to } |u_{\overline{m}}^{(i)}| = 1 \textrm{ for } i=1, \ldots, N
 \end{gathered}
\end{equation}
where $\mathrm{Tr}$ stands for $\mathrm{Trace}$. Since $Q_{\overline{m}}$ is a positive semidefinite Hermitian matrix, the optimization problem in Eq. (\ref{eq:phaseret original}) is equivalent to the optimization problem
\begin{equation}\label{eq:nonconvex phaseret}
 \begin{gathered}
  \textrm{minimize}\,\,\textrm{Tr}(Q_{\overline{m}} U_{\overline{m}}) \\
  \textrm{subject to } U_{\overline{m}} = U_{\overline{m}}^{\dagger}, \, \diag{U_{\overline{m}}} = 1, \, U_{\overline{m}} \succeq 0, \\ 
  \textrm{rank}(U_{\overline{m}}) = 1.
\end{gathered}
\end{equation}
In Eq. (\ref{eq:nonconvex phaseret}),  $U_{\overline{m}} \succeq 0$ denotes the positive semidefinite constraint on $U_{m}$. The  $\diag{U_{\overline{m}}}=1$ constraint arises from the fact that $U_{\overline{m}} = u_{\overline{m}}u_{\overline{m}}^{\dagger}$ and $|\underline{u}^{(i)}_{\overline{m}}| = 1$ implies that the diagonal elements of $U_{\overline{m}}$ will necessarily equal one. Eq. (\ref{eq:phaseret original}) and Eq. (\ref{eq:nonconvex phaseret}) constitute different formulations of equivalent optimization problems that are still difficult to solve because of the non-convex rank constraint in Eq. (\ref{eq:nonconvex phaseret}).  Dropping the non-convex rank constraint in Eq. (\ref{eq:nonconvex phaseret}) yields the complex semidefinite program
\begin{equation}
 \begin{gathered}\label{eq:phasecut sdp}
 U_{\overline{m}} = \argmin \textrm{Tr}(Q_{\overline{m}} U_{\overline{m}}) \\
   \textrm{subject to } U_{\overline{m}} = U_{\overline{m}}^{\dagger}, \,\diag{U_{\overline{m}}} = 1, \,U_{\overline{m}} \succeq 0.
 \end{gathered}
 \end{equation}
Eq. (\ref{eq:phasecut sdp}) is a convex optimization problem and can be solved efficiently using numerical solvers. We solve Eq. (\ref{eq:phasecut sdp}) using the \texttt{cvx} package \cite{grant}. If the $N \times N$ matrix $U_{\overline{m}}$ thus obtained via Eq. (\ref{eq:phasecut sdp}), has rank $1$ then we have exactly solved the original optimization problem in Eq. (\ref{eq:phaseret original}) because it is equivalent to Eq. (\ref{eq:nonconvex phaseret}). We can compute the $N \times 1$ vector  $u_{{\sf opt}}$ from $U_{\overline{m}}$ by computing the eigen-decomposition of $U_{\overline{m}}$ and setting
\begin{equation}\underline{u}_{\overline{m}} = \underline{v}_{1}(U_{\overline{m}}),\end{equation}
where $\underline{v}_{1}(U_{\overline{m}})$ is the eigenvector of $U_{\overline{m}}$ associated with its largest eigenvalue. We apply the same procedure as well when $U_{\overline{m}}$ is not rank $1$. The theoretical guaranties accompanying PhaseCut \cite{irene} establish that when $N > O(M \log M)$, then this procedure will perfectly recover $\underline{u}_{\overline{m}}$  with extremely high probability in the noise-free setting when the columns of the matrix $A$ are drawn at random as we have. We then obtain the desired $M \times 1$ vector $\underline{z}_{\overline{m}}$ from $\underline{u}_{\overline{m}}$  via Eq. (\ref{eq:zfromu}).

Having estimated ${\underline{z}}_{\overline{m}}$, from $N > O(M \log M)$ measurements, we now describe how to form a maximally focused spot at pixel $\overline{m}$.

\subsection{Focusing wavefront}\label{foc_wave}
We next determine the incident field vector $\underline{a}$ that maximizes $|b_{\overline{m}}| = |\underline{z}_{\overline{m}}^{\dagger} \underline{a}|$ for a predefined pair $\overline{m}$  subject to the energy constraint $||\underline{a}||_{2}^{2} =1$ .  We first consider the case where we have full amplitude and phase control of all $a_{m}$. This optimization problem can be expressed mathematically as:
\begin{equation}\label{eq:focusing ampphase}
\underline{a}_{{\sf opt}} = \arg \max_{||\underline{a}||_2=1}  |\underline{z}^{\dagger}_{\overline{m}} \underline{a}|.
\end{equation}
Eq. (\ref{eq:focusing ampphase}) has a closed-form solution that is given by
\begin{equation}\label{eq:aopt}
\underline{a}_{{\sf opt}}= \dfrac{\underline{z}_{\overline{m}}}{||\underline{z}_{\overline{m}}||_2}.
\end{equation}
We now consider the case where we have phase-only control of $a_{m}$, \textit{i.e.}, $|a_{m}|^{2} =1/M$. This leads to the optimization problem:
\begin{equation}\label{eq:focusing phase}
\underline{a}_{{\sf opt}} = \arg \max_{ |a_{m}|^2 =1/M,}  |\underline{z}^{\dagger}_{m} \underline{a}|.
\end{equation}
Eq. (\ref{eq:focusing phase}) has a closed-form solution that is given by
\begin{equation}\label{eq:aopt_2}
\underline{a}_{{\sf opt}}= \dfrac{1}{\sqrt{M}} \exp(i \arg[  \underline{z}_{\overline{m}}]),
\end{equation}
where
$$ \exp(i \arg[\underline{z}_{\overline{m}}]) = \begin{bmatrix} \exp\left(i \arg[ \underline{z}^{(1)}_{\overline{m}}]\right) \\ \hdots \\ \exp\left(i \arg[ \underline{z}^{(M)}_{\overline{m}]}]\right) \end{bmatrix}.$$

\section{Experimental Results}
\begin{figure}[htbp]
\begin{center}
\includegraphics[width=3.5in]{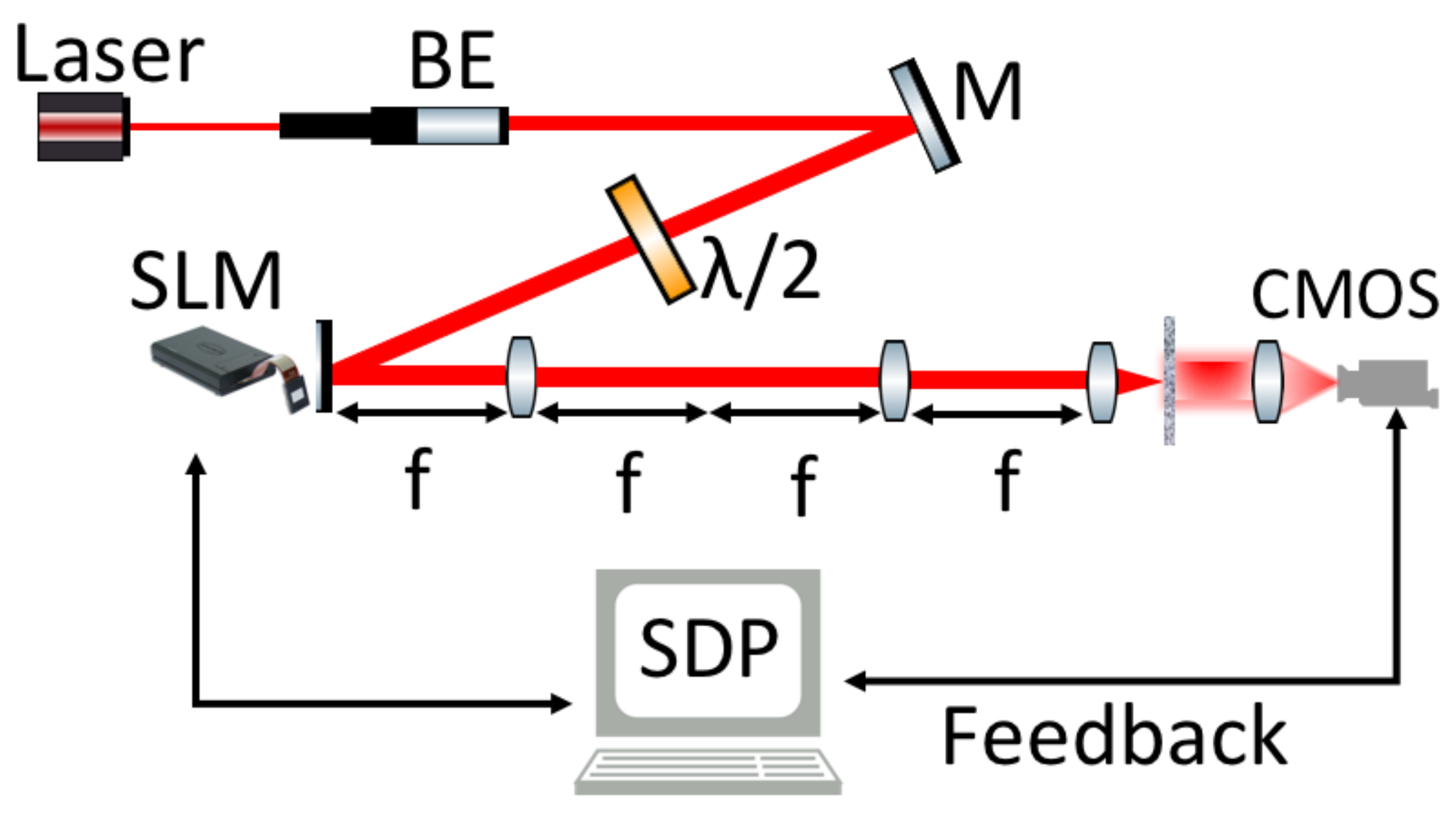}
\caption{A 633 nm CW laser is expanded then spatially filtered through a lens and a 100 $\mu$m pinhole (not shown). The polarization optics is placed before the SLM. The homogeneous beam is then reflected by the SLM that is $4f$ imaged onto the sample. The resulting speckle pattern from the scattering sample is collected by a lens and imaged on the CMOS camera.}
\label{opt_desg}
\end{center}
\end{figure}
\begin{figure}[h!]
\begin{center}
\includegraphics[width=5.0in]{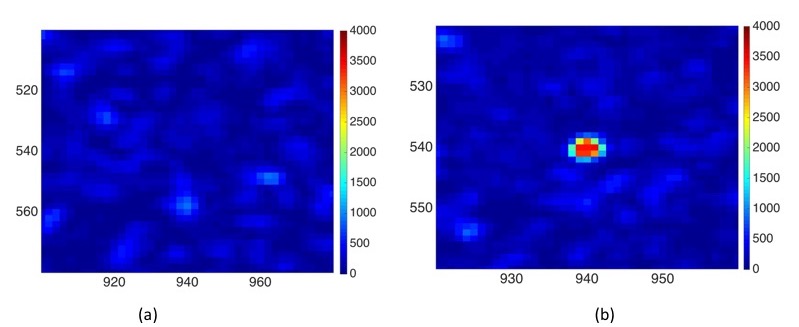}
\caption{(a) Zoomed-in image of the measured average speckle pattern. (b) High intensity focus at $(940,540)$ obtained using the optimal wavefront.}
\label{start_end_img}
\end{center}
\end{figure}
\begin{figure}[htbp]
\begin{center}
\includegraphics[width=4.0in]{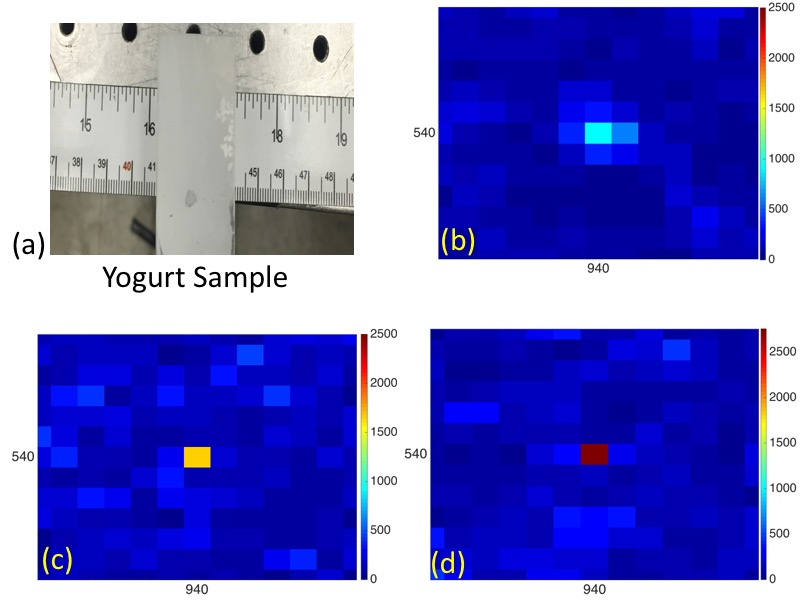}
\caption{The intensity at pixels surrounding $(940,540)$ after a focus is generated at $(940,540)$ using the SDP algorithm when b) $M = 36$, c) $M = 64$ and d) $M= 100$, respectively.}
\label{yog_samp}
\end{center}
\end{figure}
\begin{figure}[h!]
\centering
\includegraphics[width=3.0 in]{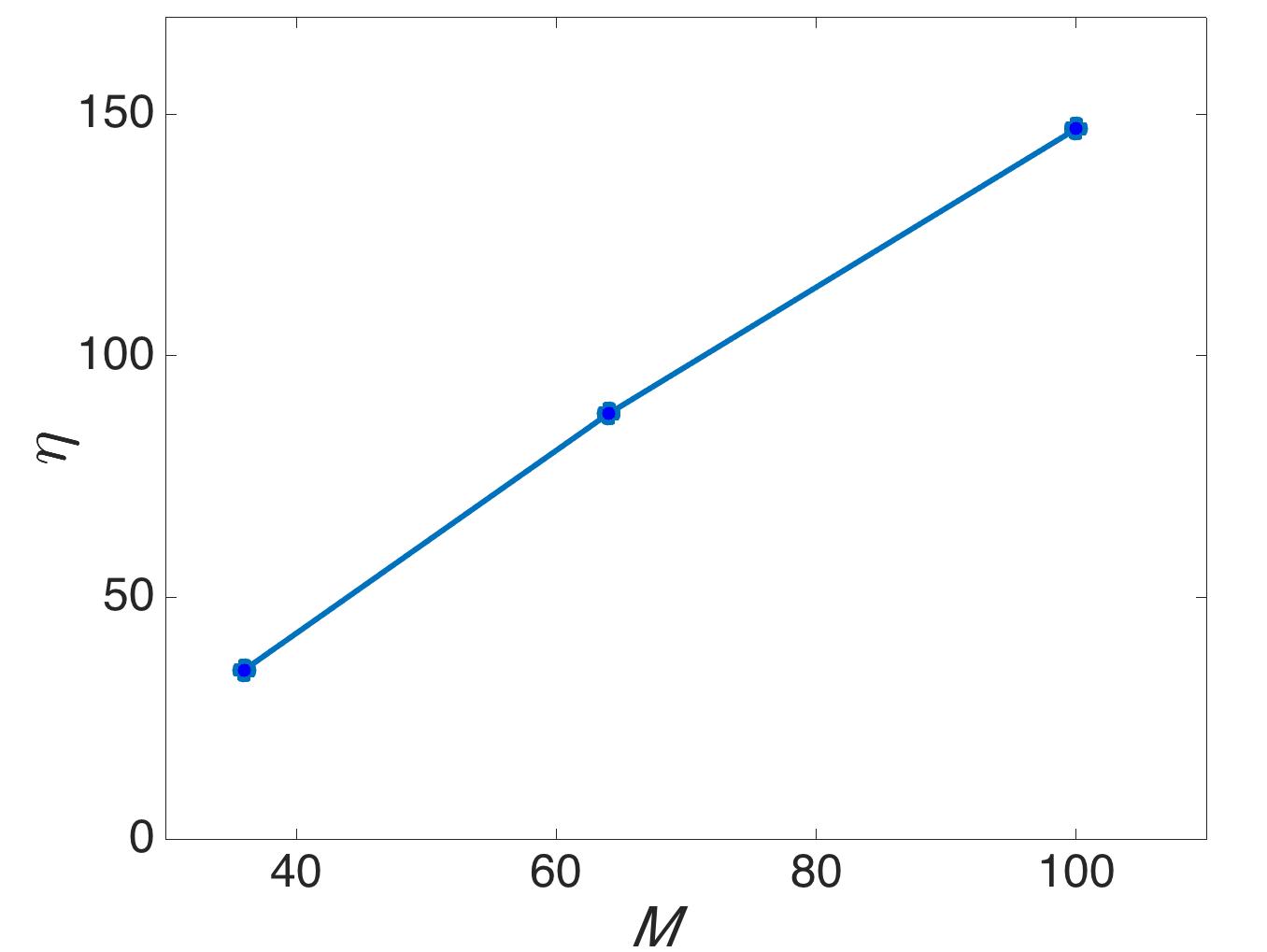}
\caption{The enhancement factor as a function of $M$ for the setup described in Fig. \ref{yog_samp}.}
\label{enhvssup}
\end{figure}
\begin{figure}[h!]
\begin{center}
\includegraphics[width=3.0 in]{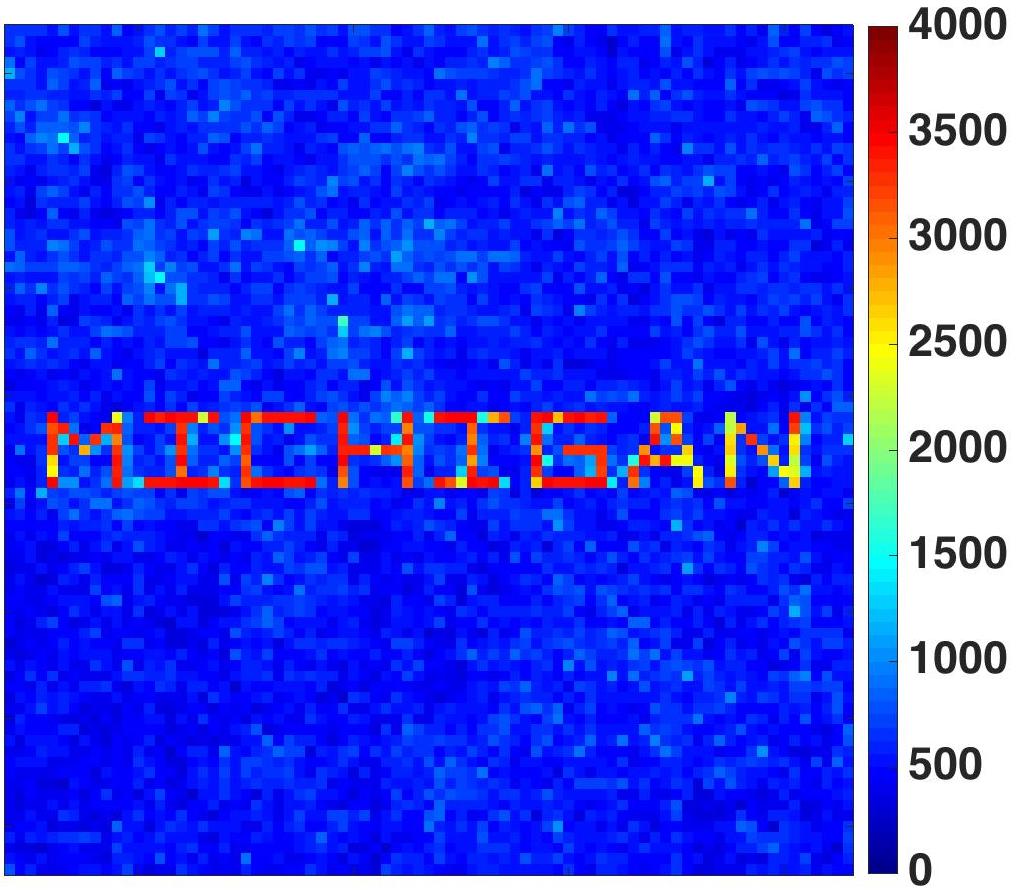}
\caption{With $M = 100$ and a ground glass as a scattering medium, we compute the optimal focusing wavefront for values of  $(\overline{k},\overline{l})$ that spell out ``MICHIGAN''. Here we plot the maximum intensity of transmitted fields for all $(\overline{k},\overline{l})$. See \url{https://vimeo.com/194525444} for an etch-a-sketch type video of the process.}
\label{michigan}
\end{center}
\end{figure}

Fig. \ref{opt_desg} shows the experimental setup. The light source is a single longitudinal mode CrystalLaser diode laser with wavelength $\lambda \approx 633$ $nm$ and output power 50 $mW$. The beam is expanded to a diameter of 20 $mm$ by a beam collimator. A set of polarization optics is used to  select the suitable polarization state of the incident beam to achieve phase only modulation. The SLM is the a phase only Holoeye PLUTO. It is a  LCOS (Liquid Crystal on Silicon) micro-display with full HD resolution (1920 x 1080 pixel) and 8 $\mu m$ pixel pitch. The surface of the SLM is $4f$ imaged and focused on the scattering sample by a lens with a 50 mm focal length. The Fourier plane of the backside of the sample is imaged onto a CMOS camera. The detector  is a PhotonFocus camera series MV1-D2048 with 2048 X 2048 resolution and pixel size 8 $\mu m$. 
\\The samples used for this experiment are ground or frosted glass and yogurt. The glass diffuser is 2$mm$ thick and 120 grit.
The yogurt sample is prepared using plain white yogurt spread in between two thin microscope slide that are pressed together to form a thin white translucent sample similar to white paint on glass. The yogurt sample thickness is $ L  \sim 150 \mu m$. The transport mean free path length ($l_s^* $) of this sample at $\lambda \approx 633$ is extracted from the angular width at half maximum of its coherent backscattering  peak: $l_s^* =  120 \mu m$. 
\\The SLM display is subdivided into $M$ equally sized squares that are dubbed superpixels. For each incident wavefront the superpixels' phases are randomly set between $[0, 2\pi]$. The corresponding transmitted intensities are measured. 
\begin{figure}[htbp]
\begin{center}
\includegraphics[width=3.5in]{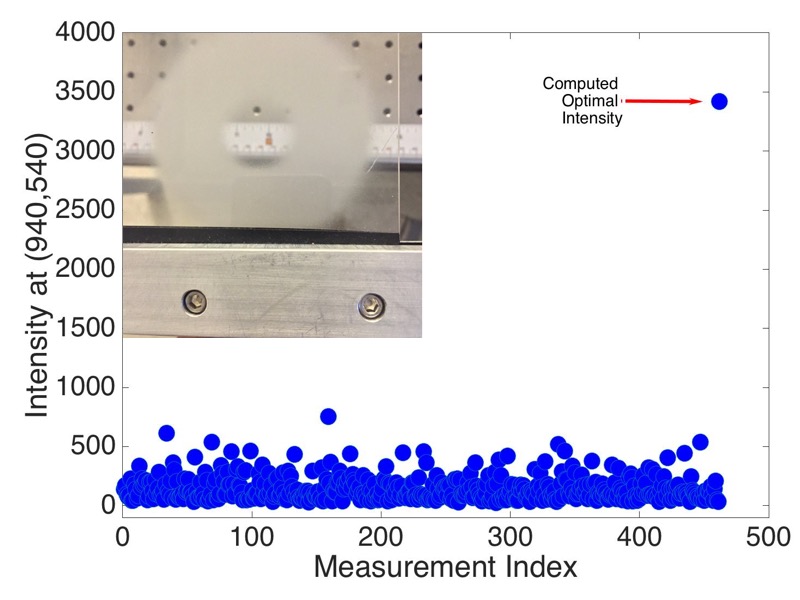}
\caption{(a) groundglass sample. (b) A plot of the intensity $|b_{(\overline{k},\overline{l})}|$ for $(\overline{k},\overline{l}) = (940,540)$ for $\lceil M\log M \rceil \propto 461$ random trial fields (here $M = 100$) followed by the intensity using the optimal wavefront $\underline{{a}}_{{\sf opt}}$ as in  Eq. (\ref{eq:aopt_2}).}
\label{comp_phase}
\end{center}
\end{figure}
Fig. \ref{comp_phase} displays $N +1$  transmitted field intensity measurements where $M = 100$ and $N = \lceil M \log M \rceil \propto 461$. Here, ground glass is used as the scattering sample shown in Fig. \ref{comp_phase}. The $461$ field intensities are measured at pixel $(\overline{k}, \overline{l})$) = $(940, 540)$  (note: in this experiment we used a different number of pixels in $A$ and $B$ as alluded to in Section \ref{prelim}; in what follows, $M$ refers to the number of pixels in $A$ and the number of pixels in $B$ is constant and much larger). The algorithm generates the incident field $\underline{a}_{{\sf opt}}$ that produces the maximum field intensity at the desired focus using Eq. (\ref{eq:aopt_2}). The $462$-th intensity measurement displayed corresponds to that produced by the optimal wavefront $\underline{a}_{{\sf opt}}$.
\\In what follows, we define the enhancement factor $\eta_{\overline{m}}$ as \cite{Ivo}: 
\begin{equation}
\eta_{\overline{m}} = \frac{|b^{{\sf opt}}_{\overline{m}}|^2}{|\overline{b}_{\overline{m}}|^2}
\end{equation}
where $b^{{\sf opt}}_{\overline{m}}$ is the intensity at $(\overline{m})$ when $\underline{\overline{a}}_{{\sf opt}}$   and $\overline{b}_{\overline{m}}$ is the average intensity at $\overline{m}$ over the $N$ training realizations. In Fig. \ref{start_end_img}(a), we show $\overline{b}_{m}$ for values of $m$ centered around $(940,540)$.   Fig. \ref{start_end_img}(b) displays the enhanced intensity at pixel $(940, 540)$; this yields an enhancement factor of $\eta_{\overline{m}} = 48$ which \textit{in par} the results of the ratio of the intensity at focus to the mean intensity of the speckle outside generated in \cite{popoff}. 
\\The SDP algorithm is stable and fast enough to be relevant for samples that are quasi-static such as a fresh yogurt sample. This sample has a speckle persistence time in the minutes. This is a compromise between the ground glass sample speckle pattern which is static for hours, and a live biological tissue which has speckle persistence time in the order of milliseconds. The measurements in Fig. \ref{yog_samp} and  Fig. \ref{enhvssup} show that for a plain yogurt sample the intensity enhancement at the desired point increases as the number of superpixels increases. Choi \textit{et al} \cite{choi} showed that once the TM is generated, the medium can be used as an unconventional lens to achieve imaging beyond the diffraction limit of the optical system. To this end, we show in Fig.  \ref{yog_samp} that the generated focus can be confined to within a single detector pixel.  
\\The SDP algorithm also allows for the generation of as many foci as desired. This is accomplished with the same initial set of $N=M\log M$  random excitations used to generate a single focus. The number of rows of the TM required now exceeds one and equals the number of foci. Fig. \ref{michigan} shows a set of foci generated beyond the scattering medium that traces out  $MICHIGAN$ comprised of 157 foci. This method becomes a relevant alternative in the deep medium regime where fields de-correlation occur and the memory effect may no longer hold.
Clearly, SDP-based wavefront shaping has the potential to be used in tandem with  fast light modulators to perform fast measurements in dynamic media.  
\\

\section*{Conclusions}
We demonstrate a new method for controlling light transmission through highly scattering random media by shaping wavefronts that optimally focus the incident beam to any position, or sequence of positions, on the far side of a scattering medium. The  optimal wavefront for a particular focus position is determined in closed-form from knowledge of  the complex-valued transmission matrix, which we determine using a sequence of intensity-only measurements, without the need for a reference beam, using a semidefinite programming based phase retrieval method. Once the transmission matrix is thus determined, we demonstrate that increasing the number of modes increases the intensity of the focus and that we can steer the focus.


\section*{Acknowledgements}
The authors would like to thank Professor Meng Cui for his generosity with time, instruments, and specially for the many informative discussions. This work was supported by a DARPA Young Faculty Award D14AP00086. The authors also thank the creators and the maintainers of the website \texttt{http://wavefrontshaping.net/} -- the many tips and suggestions and software posted there allowed for this project to come together sooner than it would have otherwise.  

\section*{Author contributions statement}

Study conceived and designed: M.N. and R.R.N., Performed the experiments: M.N., Sample preparation and chracterization: M.N and N. M. E., Algorithm developed by: R.R.N and E.M. Code generated: R.R.N, E.M, M.L. and M.N., Analyzed the data: M.N., T.B.N., and M.L. Wrote the manuscript: M.N., T.B.N., M.L, E.M., R.R.N.  The photograph of the samples:  glass diffuser insert Fig \ref{comp_phase}  and the yogurt Fig \ref{yog_samp}  were taken by MN. All authors contributed in revision.

\section*{Additional information}
\textbf{Supplementary information} accompanies this paper See \url{https://vimeo.com/194525444}
\\\textbf{Competing financial interests:} The authors declare no competing financial interests.

\end{document}